# Captions Are Worth a Thousand Words: Enhancing Product Retrieval with Pretrained Image-to-Text Models

Jason Tang*
jasontang@cs.toronto.edu
University of Toronto
Toronto, Canada

Garrin McGoldrick
Marie Al-Ghossein
Ching-Wei Chen
garrin.mcgoldrick@crossingminds.com
marie.alghossein@crossingminds.com
chingwei.chen@crossingminds.com
Crossing Minds, Inc
San Francisco, CA, USA

## ABSTRACT

This paper explores the usage of multimodal image-to-text models to enhance text-based item retrieval. We propose utilizing pretrained image captioning and tagging models, such as instruct-BLIP [6] and CLIP [24], to generate text-based product descriptions which are combined with existing text descriptions. Our work is particularly impactful for smaller eCommerce businesses who are unable to maintain the high-quality text descriptions necessary to effectively perform item retrieval for search and recommendation use cases. We evaluate the searchability of ground-truth text, image-generated text, and combinations of both texts on several subsets of Amazon's publicly available ESCI dataset [26]. The results demonstrate the dual capability of our proposed models to enhance the retrieval of existing text and generate highly-searchable standalone descriptions.

## CCS CONCEPTS

• **Information systems** → **Information extraction**.

## KEYWORDS

Information Retrieval, Search, Recommender Systems, Large Language Models, Transformers, Generative Image-to-Text, Image Tagging, Image Captioning, Computer Vision, Conversational Agents, eCommerce

---

*Work done while at Crossing Minds

## 1 INTRODUCTION

Online retail, or eCommerce, represents a rapidly growing segment of the global economy. Its share of all retail sales (including both online and brick-and-mortar) is estimated to have grown from 18% in 2017, to 35% in 2022 [11] (accelerated in part by the societal changes brought about by the COVID-19 pandemic) and is projected to reach 41% by 2027. As with traditional brick-and-mortar shopping, a online retailer's primary goal is to help their customer find and discover products that they ultimately purchase. In the context of eCommerce, the primary means for product discovery are search and recommendations[33], with major research areas including search matching [29], ranking [2], recommender systems [32], and more recently, conversational recommender systems [10].

Most search and recommendation technologies rely on knowledge of the product catalog to retrieve relevant items for a given use case. Product metadata can include title, description, attributes (such as size, color, compatibility, etc) as well as images of the product. The quality of this product metadata [18, 20] has a direct effect on the efficacy of these technologies, especially as more large-language-model-based technologies are utilized for search [14, 31] and recommendations [5, 10, 12, 35]. However, for the average eCommerce retailer, obtaining high-quality product metadata is not a trivial task. It can be dependent on the quality of the metadata provided by the manufacturer, as well as the various data sources and supply chains it needs to travel through before getting into the retailer's catalog management system. To solve this problem, many online retailers employ costly human annotators to fix or enrich their product metadata to optimize for product discovery applications. On the other hand, nearly all eCommerce websites maintain a high-quality product image catalog that caters to the fundamentally visual nature of online shopping.

Several approaches have been proposed to automatically extract product attributes from existing text attributes [17, 19, 39]. More recent approaches use image processing and computer vision techniques to predict attributes from product images as well [36, 43]. Building upon that work, this paper investigates the use of state-of-the-art pre-trained image captioning and tagging models as a means to extract textual product attributes that can enhance search and discovery capabilities on eCommerce sites, which are increasingly reliant on natural-language-based technologies.

Our contributions are as follows:

(1) We leverage image-to-text generation to enable and improve text-based item retrieval in conversational recommendation agents.
(2) We perform a literature review on current state-of-the-art multimodal image-to-text models.
(3) We analyze LLM-based conversational agents as a preprocessor for user queries.
(4) We evaluate proposed methods on the Amazon ESCI dataset [26] against several baselines and the original human-generated text.

## 2 RELATED WORK

In this study, we evaluate a range of image-to-text translation techniques. As such, our work is directly related to or adjacent to the following topics of transformer models, conversational recommendation, and image tagging and captioning systems.

### 2.1 Transformer Models

Attention-based transformer models [34] have risen to prominence across both academia and industry for their strong performance on a variety of natural language tasks. In the realm of computer vision, Vision Transformers (ViT) [9] have also found success by applying the transformer architecture directly to linear embeddings of image patches. Despite Convolutional Neural Networks (CNNs) having built-in assumptions that provide an advantage in the image domain [13], vision transformers convincingly demonstrate stronger performance on most vision tasks given ample training data.

For semantic textual similarity (STS) tasks, transformer-based models such as Bidirectional Encoder Representations (BERT) [8] demonstrated state-of-the-art performance by computing pairwise similarity scores between sentences.

In our project, transformers and ViTs are used as text and image backbones, respectively, in our multimodal image-to-text models. We also fine-tune pre-trained cross encoders and transformers to perform text-to-query similarity scoring for item retrieval.

### 2.2 Discriminative Image-to-Text Tagging

Contrastive Language-Image Pre-training (CLIP) [24] represents the cutting edge in multimodal representation learning, utilizing vision transformers [9] and text transformers [34] backbones for image and text embedding, respectively. By leveraging contrastive pre-training on the LAION-5B [30] dataset, CLIP effectively learns to project relevant images and text close together within the embedding space. Specifically, for a batch of aligned image-text pairs, CLIP optimizes a contrastive loss to minimize the distance between the embeddings of matching images and texts, and maximizes the distance otherwise. This enables CLIP to be a highly data-efficient zero-shot transfer learner that is robust to task shifts [24], relative to supervised classification models pre-trained on ImageNet [7].

In our work, we utilize Fashion CLIP (FCLIP) [3], an adaptation of the original CLIP model fine-tuned on fashion data from Farfetch, for performing metadata tagging on client data from the apparel industry. Meanwhile, CLIP is used for general image tagging and direct image-query similarity scoring in our experiments on the Amazon ESCI dataset. See Fig. 1 for a plot of PCA-reduced image and text embeddings.

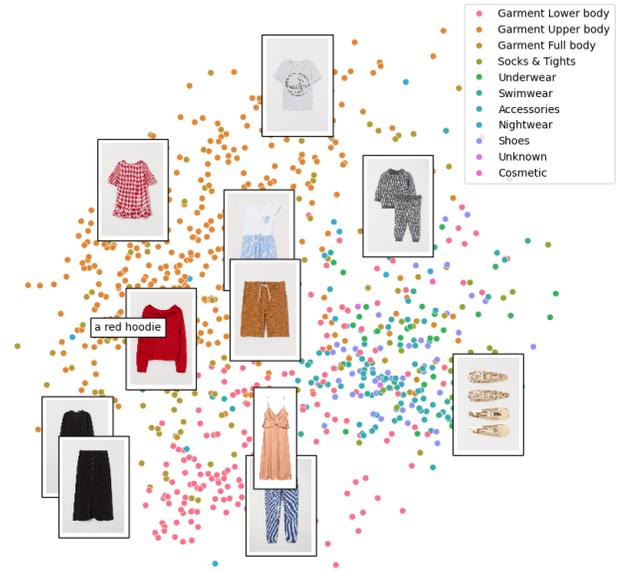

**Figure 1: PCA embeddings of image and text from a fashion dataset.**

### 2.3 Generative Image-to-Text Captioning

Bootstrapping Language-Image Pre-training (BLIP-2) [15] introduces an efficient image captioning method that only learns a lightweight transformer called a Q-Former to perform cross-modal alignment between frozen image encoders and language models. The most performant BLIP-2 models leverage vision transformers (ViT-L/14) as the image encoder and flan-t5 [25] as the text encoder. The Q-Former consists of two transformer submodules: one vision transformer interacting with the frozen ViT for visual feature extraction, and one text transformer to encode and decode text. These submodules share self-attention layers to allow cross-input interactions, and different forms of attention masking are applied depending on the pre-training objective being optimized. The authors also introduce two fine-tuning stages: one for learning queries to extract visual features that are the most informative on the text, and a second one for learning a fully connected layer to project the output representations into the frozen LLM.

InstructBLIP is another extension of BLIP-2 that reaches state-of-the-art performance on a variety of generative tasks. The main innovation is the introduction of a new Query Transformer, which, unlike the original Q-Former, enables conditional attention to ViT's visual features based on the given instructions. By further pre-training on 26 datasets in an instruction tuning format, InstructBLIP was able to outperform GPT-4, LLaVA, and MiniGPT-4 in a qualitative evaluation.

Within the computational limitations of this project, we assess the image captioning capability of these multimodal models. See Fig. 2 for an overview of both discriminative and generative models.

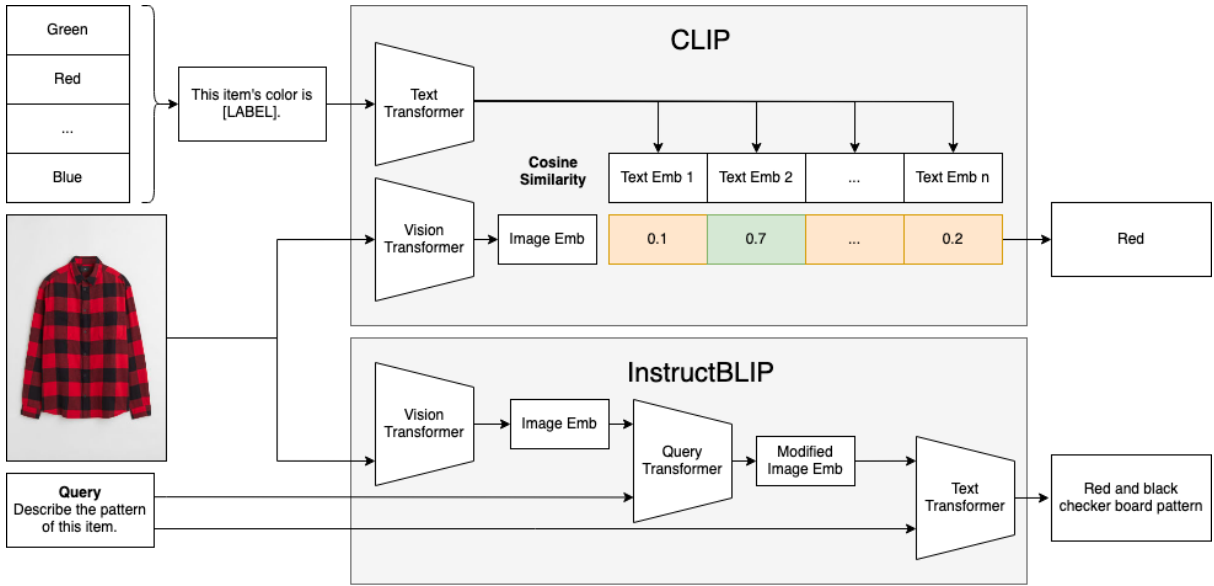

Figure 2: Overview of the Image-to-Text Tagging and Captioning models we use in experiments.

Table 1: ESCI Sample Entries

| User Query | Product Title | ESCI Label |
| --- | --- | --- |
| baby bum rash cream | Aquaphor Baby Healing Ointment - for Chapped S... | E |
| baby bum rash cream | Boudreaux's Butt Paste Diaper Rash Cream, Maxi... | S |
| baby bum rash cream | Baby Bum Brush, Original Diaper Rash Cream App... | C |
| baby bum rash cream | Diaper Rash Cream Spray by Boogie Bottoms, Tra... | I |
| invicta abalone watches for men | Invicta Men's Pro Diver Quartz Watch with Stai... | E |
| invicta abalone watches for men | Invicta Men's Pro Diver 40mm Steel and Gold To... | E |
| invicta abalone watches for men | Invicta Men's 6977 Pro Diver Collection Stainl... | S |
| invicta abalone watches for men | Invicta Pro Diver Men's Wrist Watch Stainless ... | S |

## 3 METHODOLOGIES

The objective of this project is to assess item searchability by quantifying how effectively a subset of items can be ranked according to their relevance to a given user query. We compare performance across several scenarios where item information is provided through images, textual descriptions, or a combination of both. See Fig. 3 for more details.

**TEXT - Only Text**: This reflects the original ESCI approach of comparing text to user queries. Although some descriptions may contain errors, we will assume that the original human-written item descriptions in the ESCI dataset are mostly high-quality, and we view them as the ground truth on text-query similarity search performance.

**IMG_GEN - Only Images**: We rely entirely on image-generated captions and tags as the descriptions within our text-to-query similarity search.

**TEXT+IMG_GEN - Text and Images**: We measure the ability of image-generated text to supplement existing text to improve text-to-query similarity search performance.

**IMG_DIRECT - Only Images**: We bypass the intermediary textual representation and directly embed images and queries to compute cosine similarity. Several significant limitations of this approach are the requirement of costly vector databases to store embeddings and the need for a model inference pass to embed the incoming user query for each incoming recommendation call. Without extensive optimizations, this could lead to prohibitively high latencies or costs. Furthermore, omitting the intermediate text stage forfeits the opportunity to automate or bootstrap the manual captioning process for new products.

### 3.1 Similarity Search

Our methods center on text-based query search for item retrieval within the context of eCommerce search and recommendation scenarios. We integrate our proposed multimodal methods with fine-tuned cross encoders and transformers to achieve efficient text-to-query similarity scoring and item retrieval. Additionally, we also explore the direct approach of comparing item image and user query embeddings [14]. Refer to Fig. 4 for an overview.

### 3.2 Image-to-text Models

We examine the zero-shot capabilities of generative models in generating unstructured item catalog descriptions and detecting specific features from images. Due to computational limitations, we were limited to exploring BLIP-2 and InstructBLIP with flan-t5 backbones. These captioning models were loaded across two NVIDIA 1080Ti GPUs using 16-bit floating-point half precision.

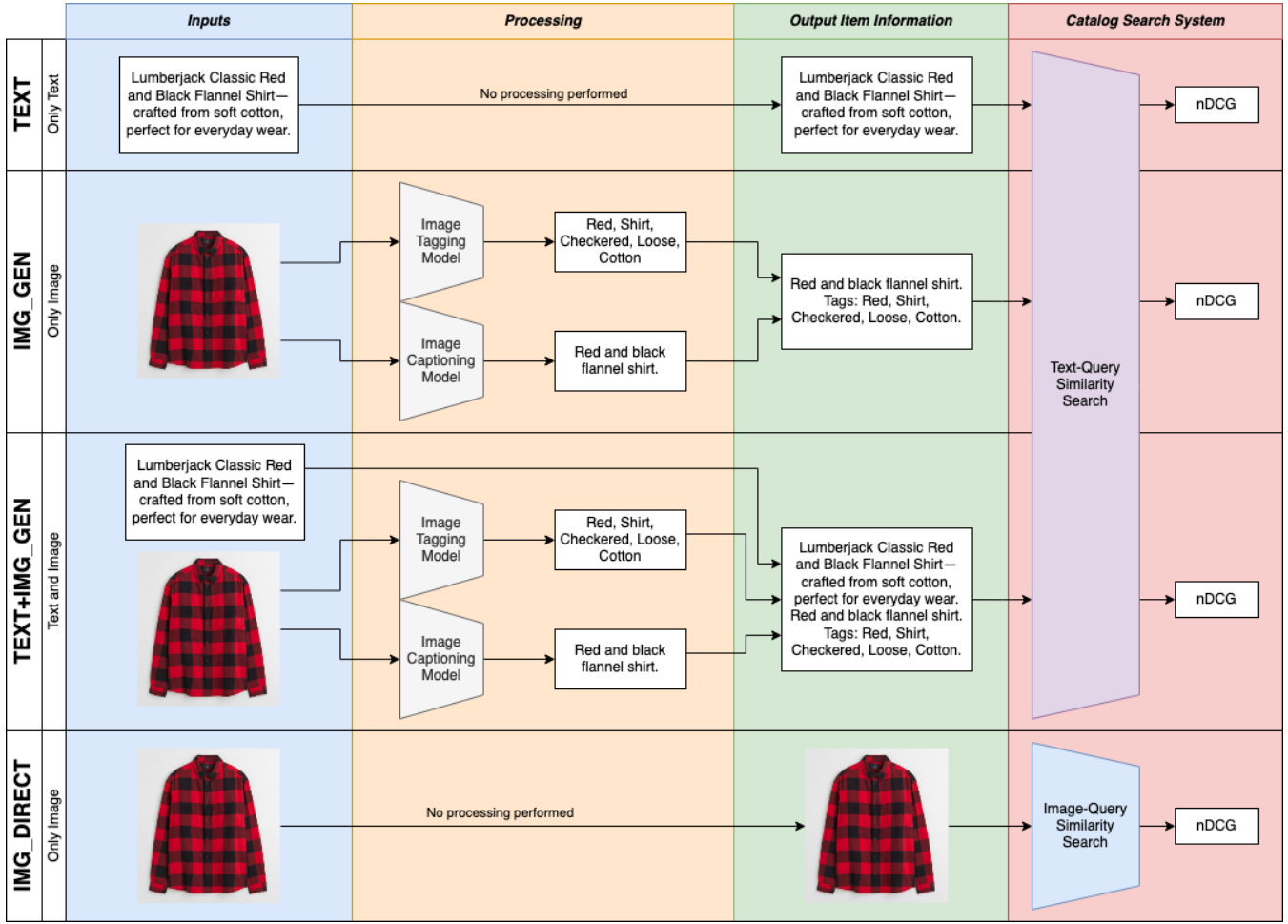

**Figure 3: Overview of our approaches on settings with different types of available data. More details about the Catalog Search Systems can be found in Fig. 4.**

We also assess the zero-shot classification abilities of discriminative multimodal models to classify metadata tags. We chose to use the broadly applicable CLIP on the Amazon ESCI dataset.

To broaden the scope of generated text, we prompt models to extract various features for each item that are likely targets of user searches, such as color, material, usage, and intended user. Additionally, we experiment with a series of differently worded prompts and compare outputs to select the best phrasing for model performance.

### 3.3 Query Preprocessing

We observe that user queries are often misspelled, too vague, or contain non-English terms. To address this, we leverage the commonsense reasoning and extensive world knowledge of ChatGPT-3.5 to refine and elaborate upon the initial user queries. An effective prompt for ChatGPT-3.5 to perform this preprocessing was: "Extract at least 5 related tags or usage keywords from queries. Output in English as a comma separated list.". See Table. 2 for examples.

### 3.4 Experiments

Our experiments investigate the efficacy of image-to-text generation in enhancing catalog search across contexts that employ images, textual descriptions, or a combination of both. Across our experiments, all pre-trained BLIP, CLIP, and transformer models used in our experiments were sourced from HuggingFace [38][1]. We also obtained an implementation of Cross Encoders from Sentence

---
[1] https://huggingface.co/

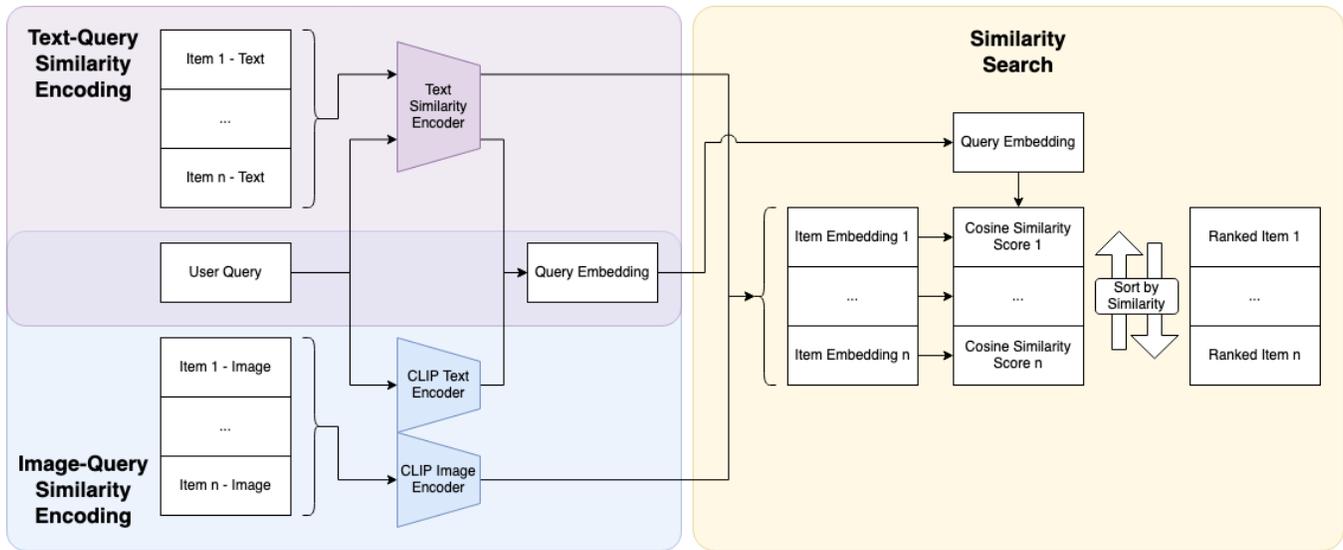

Figure 4: Our pipelines for Text-Query and Image-Query similarity search.

Table 2: ChatGPT-3.5 Preprocessing

| User Query | Processed Query |
|---|---|
| !awnmower tires without rims | lawnmower, tires, without rims |
| #20 paper bags without handle | paper bags, without handle, packaging, eco-friendly, retail |
| paws | animal, pets, claws, dogs, cats |
| apple iphone 11 pro unlocked | apple, iPhone, 11 pro, unlocked |
| 자전거트레일러 | bicycle trailer, bike trailer, cycling trailer, bike cart, bike carrier |
| 眼镜框 | eyeglass frames, glasses frames, eyewear, spectacle frames, glasses |

Transformers [27][2], and all model modifications were written using PyTorch [22][3].

Training and evaluation were conducted on Crossing Minds' internal compute cluster, which is equipped with multiple NVIDIA 1080Ti GPUs.

3.4.1 *Datasets.* Our experiments evaluate the utility of our image-generated features on the publicly available Amazon ESCI dataset [26], which features a wide range of eCommerce product categories such as entertainment, apparel, technology, and home decor (see Table 1 for examples). We focus on the English subset of Task 1 in the ESCI dataset, where a subset of Amazon products must be ranked in order of relevancy to a given user query. Each query-product pair is manually classified with a relevance label. See Table 3 for more details.

[2]https://www.sbert.net/
[3]https://pytorch.org/

Through text-query sentence similarity search, we are able to compare the retrieval capabilities between our image-generated descriptions and the original human-crafted descriptions. Within this dataset, each user query has up to 40 example products that need to be ranked, with an average of 20.3 example products per query. We will call this number of examples per query ratio E/Q for brevity. Due to the project timeline and computational limitations, our research only considers a subset of approximately 20,000 of the most popular products out of the original 482,000 unique Amazon products in the dataset. We measured product popularity based on the number of occurrences as an example for a query within the dataset.

Given that images were not included in the ESCI dataset, we used the provided product identifiers to scrape the associated Amazon product pages for images. To streamline this process, we limited our scope to the first image of each item and selected the image size with as close to a resolution of 384x384 as possible. With an average of 15–30 seconds per image retrieval, scraping the entire original catalog of 482,000 products would have taken in excess of three months to complete.

3.4.2 *ESCI Label Distribution.* See Table 4 for an overview of label distributions.

**Dataset 1: Most Popular Items ($PadSize = 0$)**

Due to the aforementioned limitations, we were only able to consider a subset of the original 482,000 products. To improve sample efficiency, we filter the dataset to only retain items that appear in a minimum of three queries as examples and eliminate all queries with no examples left. This produces a dataset with a total of 21,627 items and 19,825 queries. Additionally, we see a significant drop in the number of examples per query (E/Q) from 20.3 to 4.8.

Both the original ESCI dataset and this subset consist mainly of query-product pairs that are exact matches and substitutes, with a smaller proportion of irrelevant pairs. This label distribution indicates that this dataset primarily assesses an approach's ability to generate text that can discern between exact matches and substitutes, with a secondary emphasis on discriminating irrelevant from relevant matches. To perform this task, minute details in the item description are crucial to determining if a query-item pair is an exact match or a substitute. As such, we expect our image-generated text to perform slightly poorer in this task, given that critical information, such as dimensions, product version, and technical specifications, are not typically discernible from images alone.

Table 3: ESCI Label Definitions

| ESCI Label | Definition | User Query | Item |
|---|---|---|---|
| Exact (E) | Satisfies all constraints | Loose Fit Red Dress Shirt | Loose Fit Red Dress Shirt |
| Substitute (S) | Alternative substitute | Loose Fit Red Dress Shirt | White Dress Shirt |
| Complement (C) | Complements the desired item | Loose Fit Red Dress Shirt | Black Tie |
| Irrelevant (I) | Everything else | Loose Fit Red Dress Shirt | Water Bottle |

We note that a major limitation with this dataset is the small E/Q ratio of 4.8. With such a small number of items that need to be ranked, even our random baseline is able to achieve strong nDCG scores (see Fig. 5).

**Dataset 2: Most Popular Items + Random Irrelevant Samples ($PadSize \in [5, 10, 20]$)**

To mitigate this issue, we introduce a modified version of Dataset 1 where we fill each query with randomly sampled items until the query has at least $PadSize$ items. All randomly sampled items are labeled as irrelevant, since we assume that the vast majority of catalog items are irrelevant to any individual query. Moreover, considering the high proportion of relevant items in the initial dataset, it is improbable that there are a significant number of unselected relevant items remaining in the dataset.

Through this padding strategy, we are able to bring our E/Q ratio up to 20.2 when $PadSize = 20$, which aligns closely with the E/Q of 20.3 in the original ESCI dataset. The resulting label distribution better reflects the expected label distributions for item retrieval from real store catalogs, where irrelevant items substantially outnumber exactly matching and substitute items. As such, we can interpret this scenario as assessing our models more on their ability to discern relevant items from a larger pool of irrelevant items.

*3.4.3 Metrics.* We quantify the item retrieval and ranking ability of our various approaches through Normalized Discounted Cumulative Gain (nDCG). nDCG evaluates how close the first $k$ predicted ranking items are to the ideal ranking order, yielding a number from 0 to 1, with higher values indicating better performance. In our experiments, we set $k$ to the maximum length of sequences to consider all ranked items.

$$\text{DCG}_k = \sum_{i=1}^{k} \frac{\text{relevance}_i}{\log_2(i+1)}$$

$$\text{IDCG}_k = \sum_{i=1}^{k} \frac{\text{relevance}_i^{ideal}}{\log_2(i+1)}$$

$$\text{nDCG}_k = \frac{\text{DCG}_k}{\text{IDCG}_k}$$

## 4 RESULTS AND DISCUSSIONS

### 4.1 Baselines

In previous settings without existing text, we would employ methods such as random or most popular ranking that do not rely on item information. We explore several popularity scoring variations in Fig. 5.

**Random Baseline:** An initial lower-bound baseline was established by randomly ranking products. To minimize the influence of outliers, we take the median nDCG score across five random orderings.

**Most Popular Baseline:** We also establish a popularity-based baseline by assigning scores to products according to their ESCI relevance labels and aggregating across all occurrences. The most effective label scoring method was scoring $[E, S, C, I]$ labels with $[1, 0, 0, 0]$, respectively. This represents the number of times each item appears as an exact match in the training data. We also considered alternate popularity scoring methods such as the number of non-Irrelevant matches $[1, 1, 1, 0]$, and decreasing scores $[1.0, 0.1, 0.01, 0.0]$.

Our experiments reveal that both baselines achieve high nDCG scores in Dataset 1 due to the limited number of items that need to be ranked. However, as we increase the E/Q ratio with more random padding, both baselines experience up to a 40% decline in performance. As anticipated, since the most popular baseline incorporates information from training data statistics, we see our most popular baseline marginally outperform the random baseline across all settings.

Table 4: ESCI Label Distributions

| | Dataset 0 | Dataset 1 | Dataset 2 | | |
|---|---|---|---|---|---|
| Label | $PadSize = 0$ | $PadSize = 0$ | $PadSize = 5$ | $PadSize = 10$ | $PadSize = 20$ |
| E | 43.9% (79708) | 37.7% (10751) | 27.0% (10751) | 16.9% (10751) | 8.9% (10751) |
| S | 34.9% (63563) | 37.0% (10548) | 26.5% (10548) | 16.6% (10548) | 8.8% (10548) |
| C | 4.5% (8099) | 3.5% (989) | 2.5% (989) | 1.5% (989) | 0.8% (989) |
| I | 16.7% (30331) | 21.8% (6212) | 43.9% (17478) | 64.9% (41365) | 81.4% (97398) |
| # Examples | 181701 | 28500 | 39766 | 63653 | 119686 |
| # Queries | 8956 | 5935 | 5935 | 5935 | 5935 |
| E/Q Ratio | 20.3 | 4.8 | 6.7 | 10.7 | 20.2 |

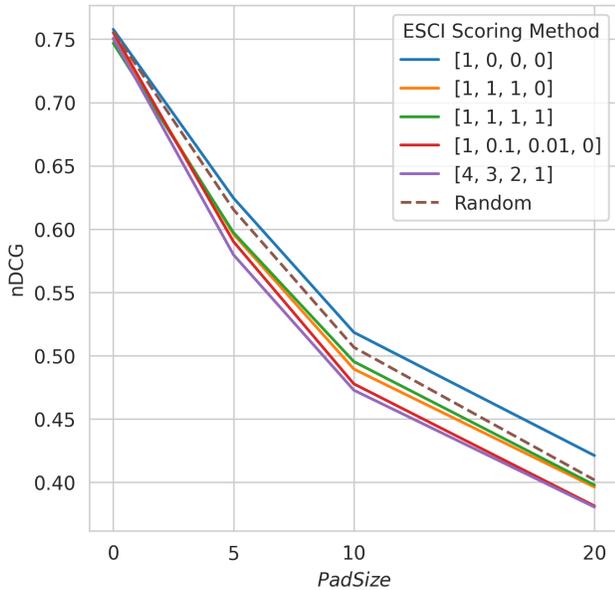

Figure 5: Ranking performance of our random baseline and several variations of the most popular baseline across different *PadSize*'s.

## 4.2 Approaches using Item Text or Images

Fig. 6 illustrates the average nDCG performance and includes error bars representing the range between the minimum and maximum nDCG values observed over four independent runs. Additionally, Fig. 7 compares these results to established baseline measures.

As anticipated, the original Amazon text outperforms our purely image-generated text, although the latter only marginally underperforms the original text and noticeably outperforms the baseline methods as *PadSize* and E/Q increase. Notably, we observe a graceful degradation of performance across non-baseline methods as the number of examples that need to be ranked grows.

Additionally, we see that augmenting the original Amazon product descriptions with our image-generated text enhances overall performance. This suggests that our image-generated text possesses the ability to capture details that the existing text may have missed. We also note that the unstructured InstructBLIP outputs are considerably more effective than CLIP-based tagging. This disparity in performance may be a result of CLIP outputs being constrained to pre-selected tag options, which could potentially be mitigated by scaling and improving the tag options list to align more closely with user search terms.

Our findings also show that the direct image-query similarity approach yields slight improvements over our other image-only methods across all datasets. However, we argue that the previously discussed drawbacks of this approach outweigh these modest gains in the context of real-world applications.

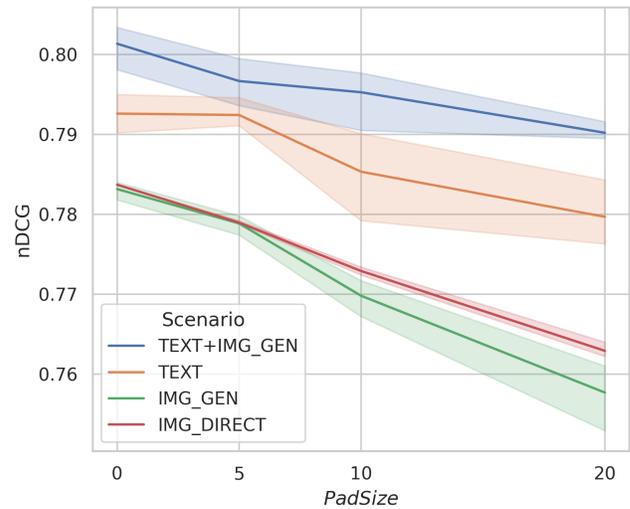

Figure 6: Ranking performance of our proposed multimodal approaches across different *PadSize*'s, highlighting the influence of different input modalities.

In summary, our experiments demonstrate the capability of our multimodal models to both supplement the searchability of existing text and generate high-performance standalone descriptions.

## 4.3 Cross Encoder versus Transformer Similarity

In our text-query search step, we evaluate two sentence similarity models: a conventional transformer model fine-tuned for semantic search task [4], and a cross-encoder transformer [28][5]. Although both models demonstrate comparable performance on datasets with smaller *PadSize*'s (see Table 5), the cross-encoder model begins to exhibit superior performance compared to the standard transformer when handling larger padding sizes. This improvement can likely be attributed to the cross-encoder's previously discussed specialized training methodology for learning semantic similarity tasks.

Table 5: Similarity Methods Comparison Across *PadSize*'s

| PadSize | Transformer | Cross Encoder |
|---|---|---|
| 0 | **0.782** | 0.781 |
| 5 | **0.775** | 0.773 |
| 10 | 0.759 | **0.764** |
| 20 | 0.740 | **0.750** |

## 4.4 GPT Preprocessing

From our experiments (Table 6), we observe that refining and elaborating upon user queries through ChatGPT-3.5 results in a slight yet consistent performance improvement across most experimental settings. This result aligns with our expectations, as our investigations identified the presence of errors, misspellings, and non-English terms within the original user queries.

---
[4]https://huggingface.co/sentence-transformers/multi-qa-mpnet-base-dot-v1
[5]https://huggingface.co/cross-encoder/ms-marco-MiniLM-L-2-v2

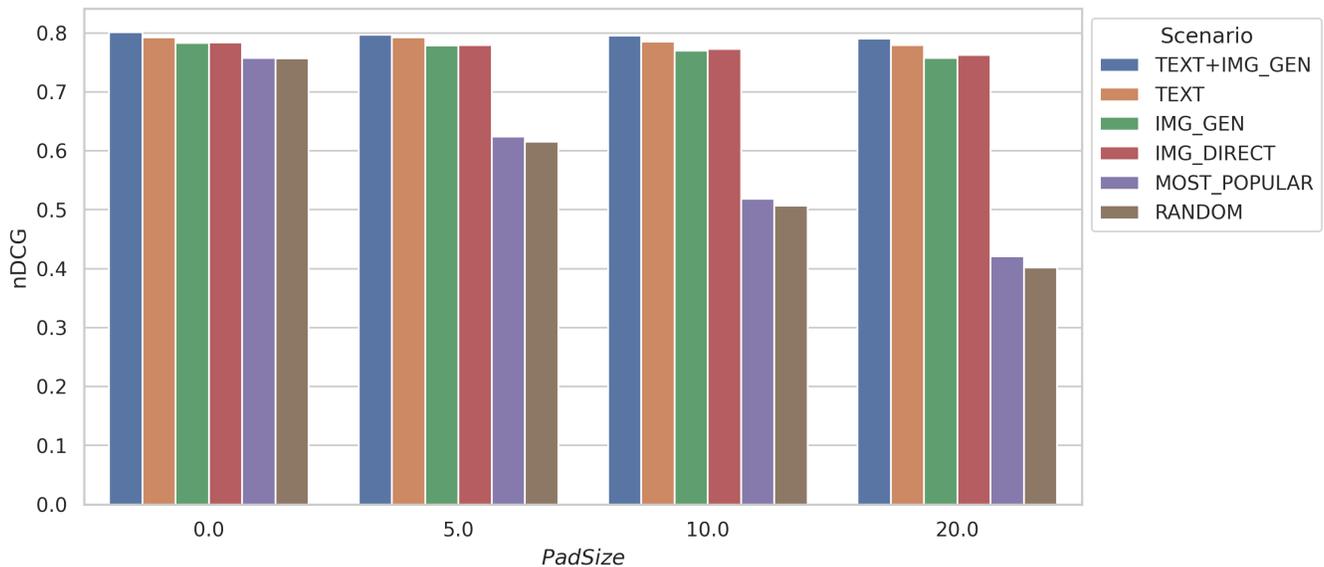

**Figure 7: Comparing the ranking performance of our multimodal methods against our baselines across different *PadSize*'s.**

**Table 6: GPT Preprocessing Comparison Across *PadSize*'s**

| PadSize | Original Query | GPT Preprocessing |
|---|---|---|
| 0 | 0.780 | **0.782** |
| 5 | 0.767 | **0.774** |
| 10 | 0.756 | **0.762** |
| 20 | 0.734 | **0.745** |

## 5 CONCLUSIONS AND FUTURE RESEARCH

In this paper, we leveraged image-to-text captioning and tagging models to extract textual information from images and enhance item retrieval. Our investigation involved two main approaches: unstructured text synthesis using the InstructBLIP [6] model with a flan-t5 [25] LLM backbone, and metadata tagging via CLIP [24] models.

Although a comprehensive evaluation of the ESCI dataset [26] was beyond the scope of this project, we introduced a sample-efficient padding method to create datasets that better reflect real-world store catalogs, where the majority of items are irrelevant to any given query. This allowed us to demonstrate the ability of our multimodal models to generate high-performing image-derived descriptions that enable eCommerce platforms without substantial text descriptions to utilize conversational agents, while also supplementing existing high-quality text to improve item retrieval performance.

### 5.1 Future Directions

*5.1.1 Large-Scale Experiments.* One direct extension involves scaling up computational resources to experiment with miniGPT-4 [42] and Vicuna [4] LLMs, which show potential advantages in long form generation compared to flan-t5 LLMs [6, 25]. With the anticipated introduction of state-of-the-art multimodal functionalities in GPT-4 [21], image captioning could be further improved beyond the capabilities of our models.

Additionally, with sufficient time or computational resources, the methods proposed in this project can be evaluated on the entire ESCI dataset, such that their abilities to discern between exact and substitute matches will be more accurately assessed.

*5.1.2 Techniques from KDD Cup 2022 ESCI Challenge.* In the KDD Cup 2022 ESCI Challenge, a variety of techniques led to an increase in the baseline nDCG score from 0.83 to 0.90. These included self-distillation [40], data augmentation, post-processing, adversarial training [1], prompt tuning, and keyword extraction [16, 23, 41]. The application of self-distillation using ensembles and soft labels effectively improved robustness to noisy data [16, 41]. Moreover, to mitigate the issue of overly simplistic queries, queries were extended using Term Frequency Inverse Document Frequency (tf-idf) [37] keywords from relevant (E, S, and C labeled) item descriptions from the training set [16, 41]. The incorporation of these techniques shows strong potential for boosting the performance of the methods outlined in this project.

*5.1.3 Evaluation under ElasticSearch.* Another extension is to assess the utility of image-generated captions for text-query searches using ElasticSearch — a popular search engine offering rapid text search capabilities and cost effectiveness. The use of systems like ElasticSearch is particularly pertinent to any real-world applications, where high performance and scalability of infrastructure are critical. We chose to use text-query cosine similarity search in our work as a preliminary metric for measuring searchability and anticipate that analogous results can be achieved in ElasticSearch-based systems.

## ACKNOWLEDGEMENTS

The team at Crossing Minds has my deepest gratitude for their unwavering guidance and support during my internship. Their invaluable insights and contributions were integral to the success of this project. I am thankful for the opportunity to have been a part of Crossing Minds.

I would like to thank my industry supervisor, Garrin McGoldrick, for overseeing my day-to-day progress at Crossing Minds. His skill in aligning my personal interests, business needs, and research requirements significantly contributed to a great internship experience.

Special thanks to my research lead, Marie Al Ghossein, for guiding my research directions and helping ideate creative solutions. This project would not be successful without her mentorship.

I would also like to thank my academic supervisor, Eldan Cohen, for his deeply useful knowledge in state-of-the-art machine learning literature and for providing critical feedback on academic writing.

Last but not least, this internship would not have been possible without the funding from the Mitacs Accelerate program.